\documentclass[twocolumn,pra,aps,amssymb]{revtex4}
\usepackage{algorithm2e}
\usepackage{amsmath,amssymb,amsfonts,amsthm}
\usepackage{multirow}
\usepackage{verbatim}
\usepackage{url}
\usepackage{comment}
\usepackage{mathtools}
\usepackage{epsf,pgf,graphicx}
\usepackage{color}
\usepackage{tikz,pgf}

\begin{document}
\newcommand{\ket}[1]{\ensuremath{\left|#1\right\rangle}}
\newcommand{\bra}[1]{\ensuremath{\left\langle#1\right|}}
\newcommand\floor[1]{\lfloor#1\rfloor}
\newcommand\ceil[1]{\lceil#1\rceil}
\newtheorem{definition}{Definition} 
\newtheorem{proposition}{Proposition} 
\newtheorem{theorem}{Theorem} 
\newtheorem{claim}{Claim}
\newtheorem{lemma}{Lemma}
\title{Device Independent Quantum Private Query with Finite Number of Entangled Qubits}
\author{Jyotirmoy Basak$^1$, Bappaditya Ghosh$^1$, Arpita Maitra$^2$ and Goutam Paul$^1$}
\affiliation{
$^1$Indian Statistical Institute, Kolkata,\\
Email: \{bjyotirmoy.93,bappaditya.ghosh86\}@gmail.com,\\goutam.paul@isical.ac.in\\
$^2$Indian Institute of Technology Kharagpur, India,\\
Email: arpita76b@gmail.com
}

\begin{abstract}
In a recent work by Maitra et al. (Phys. Rev. A, 2017), it was shown
that the existing Quantum Private Query (QPQ) protocols fail to maintain the database security if the entangled states shared between Alice and Bob are not of a certain form. So it is necessary to certify the states a priori. In this regard, the local CHSH test was proposed. However, the proposed scheme works perfectly for the asymptotic case when we have infinite number of qubits. In this brief report, we upgrade the protocol for finite number of qubits and connect the sample size to the success probability of CHSH test. We also perform a rigorous security analysis of the proposed protocol.
\end{abstract}
\maketitle

\section{Introduction}
Quantum private query is a two party mistrustful cryptographic primitive. In QPQ one of the two legitimate party, say Bob, owns a database. His job is to protect the entire database from the client's (Alice's) knowledge along with providing the element asked by the client. On other hand, the client's motivation is to extract more elements from the database beside her query.    

Giovannetti et al.~\cite{GLM} first proposed the idea of QPQ protocol followed by~\cite{GLM11,Olejnik}. However, their protocols were found difficult to implement in practice. With the motivation for practical implementation Jakobi et al.~\cite{jakobi} came out with a QPQ proposal which is based on SARG04 quantum key distribution protocol~\cite{SARG}. In 2012, Gao et al.~\cite{Gao} presented flexible QPQ protocol. Here, flexible means Bob can regulate the information of Alice about the shared key between themselves by controlling some parameters. In 2014, Yang et al.~\cite{Yang} proposed the entanglement version of~\cite{Gao} with the help of B92 QKD protocol~\cite{b92}.

Any QPQ protocol deals with database security and user privacy. Database security guarantees that the protocol never leaks any element of the database to the client except the query he or she has made. On the other hand, user privacy prevents the database owner to know the query of the client. Very recently, Maitra et al.~\cite{Maitra} showed that if the entangled states shared between Bob and Alice are not in a specific form, then Alice can exploit a strategy by which she can extract more information than what is suggested by the protocol. In other words, if the entangled states are not in a certain form, then the database security becomes vulnerable. To resist such attack, they proposed local CHSH test~\cite{Lim}. Observing the outcomes of the test, they certify whether the states are suitable for further use in QPQ protocol. 

For DI protocols~\cite{acin06a,acin06b,scarani06,Acin} involving maximally entangled state, e.g., for QKD-type protocols, the success probability of Bob has to be precisely the maximum possible, i.e., $cos^2 \pi/8 \approx 0.85$ and hence infinitely large number of qubits are required to estimate this probability. However, for the QPQ protocol, for any given $\theta$, the maximum probability may be less than $cos^2 \pi/8$. Thus, we can reduce the number of qubits at the cost of allowing the estimated probability to deviate from the expected maximum probability by a negligible amount. 

Due to the above deviation, we have to allow some information leakage to Alice. However, we show that this information leakage entirely depends on how much deviation we should accept for the test. The order of the information leakage is exactly same as the order of the deviation allowed. 

In this report we follow up the work of~\cite{Maitra} in the motivation towards bridging the gap between theory and practice.

\section{Modification Towards Finite Number of Entangled States}
\label{sec}
In~\cite{Maitra}, it has been shown that if the entangled states shared between Bob and Alice are not in a certain form, then Alice can exploit a cleaver strategy to gain more information about the shared key than what is suggested by the protocol. Thus, to ensure the security of the QPQ protocol, it is necessary to test the shared states a priori. As Alice is considered as an adversary here, so it is Bob who tests the states. The test proposed in~\cite{Maitra} is actually the CHSH test performed locally with non-maximally entangled states.
Under some reasonable assumptions the scheme presented in~\cite{Maitra} works perfectly when the numbers of the entangled states tend to infinity.

In this section, we will discuss how we can modify the protocol~\cite{Maitra} for the finite numbers of entangled states. Firstly, we try to estimate the number of qubits for the local CHSH test in an optimal way. By the word `optimal' we want to mean that the estimated value should be the minimum number of samples required for the local CHSH test. Secondly, we show that the deviation which we have to allow to upgrade the protocol for finite sample size does not open a security loop-hole in the protocol. 

\subsection{Maximization of success probability} 
\label{max}
In DI-QPQ protocol~\cite{Maitra}, Bob and Alice share entangled states of the form $\frac{1}{\sqrt{2}}(|0\rangle_{B}|\phi_0\rangle_{A}+ |1\rangle_{B}|\phi_1\rangle_{A})$, where, $|\phi_{0}\rangle_{A}=\cos{(\frac{\theta}{2})}|0\rangle + \sin{(\frac{\theta}{2})}|1\rangle$ and $|\phi_{1}\rangle_{A}=\cos{(\frac{\theta}{2})}|0\rangle -\sin{(\frac{\theta}{2})}|1\rangle$. The value of $\theta$ is known to all. Bob chooses two measurement bases namely $\{\ket{\psi_1},\ket{\psi_1^{\perp}}\}$ and $\{\ket{\psi_2},\ket{\psi_2^{\perp}}\}$, to play the local CHSH game. Here, $\ket{\psi_1}=\cos\frac{\psi_1}{2}\ket{0}+\sin\frac{\psi_1}{2}\ket{1}$
and $\ket{\psi_2}=\cos\frac{\psi_2}{2}\ket{0}+\sin\frac{\psi_2}{2}\ket{1}$. Now, for a particular value of the angles $\psi_1$, $\psi_2$ and $\theta$, only Bob can calculate the success probability value of the local CHSH game, hence, preventing Alice to manipulate the states and the measurement devices. Here we propose a modification of the scheme~\cite{Maitra}.

In the DI-QPQ protocol~\cite{Maitra}, Bob gets the success probability in terms of $\theta$, $\psi_1$ and $\psi_2$ which is equal to $\frac{1}{8}(\sin{\theta}(\sin\psi_1+\sin\psi_2)+\cos\psi_1-\cos\psi_2)+\frac{1}{2}$. To maximize the quantity, we have to maximize $\sin{\theta}(\sin\psi_1+\sin\psi_2)+\cos\psi_1-\cos
\psi_2$. 

Now, we can write,
\begin{eqnarray*}
&&\sin{\theta}(\sin\psi_1+\sin\psi_2)+\cos\psi_1-\cos\psi_2\\
&=&\sin{\theta}\sin\psi_1+\sin{\theta}\sin\psi_2+\cos\psi_1-\cos\psi_2 \\
&=&(\sin{\theta}\sin\psi_1+\cos\psi_1)+(\sin{\theta}\sin\psi_2-\cos\psi_2)
\end{eqnarray*}

Setting $\sin{\theta} = r \cos{\phi}$ and $1 = r \sin{\phi}$, we get
\begin{eqnarray*}
& & (r\cos{\phi}\sin\psi_1+ r \sin{\phi}\cos\psi_1)\\
& & +(r \cos{\phi}\sin\psi_2- r \sin{\phi}\cos\psi_2) \\
&=& r (\sin(\psi_1 + \phi) + \sin(\psi_2 - \phi) ), 
\end{eqnarray*}
where $r^2=1+\sin^2{\theta}$, $r \cos{\phi}=\sin{\theta}$ and $r \sin{\phi}=1$. Thus we get, $\tan{\phi}=\mathrm{cosec}~{\theta}~$ i.e, $\phi=\tan^{-1}(\mathrm{cosec}~{\theta}~)$.

Again, the value $r ( \sin(\psi_1 + \phi) + \sin(\psi_2 - \phi) )$ will be maximum when both $\sin(\psi_1 + \phi)=1$ and $\sin(\psi_2 - \phi)=1$ i.e, when $(\psi_1 + \phi)=\frac{\pi}{2}$ and $(\psi_2 - \phi)=\frac{\pi}{2}$. From that we get, $\psi_1=(\frac{\pi}{2}- \phi)$ and $\psi_2=(\frac{\pi}{2}+ \phi)$.

\begin{figure}[htbp]
\includegraphics[width=0.45\textwidth]{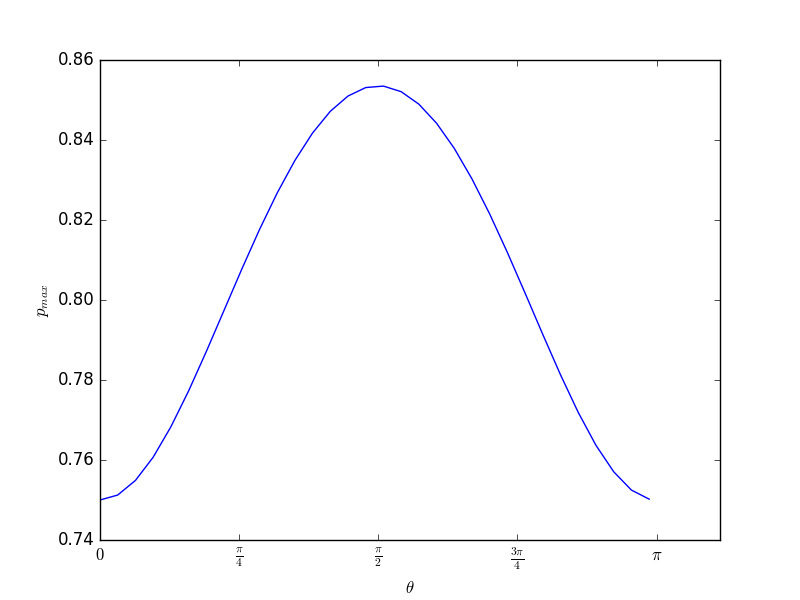}
\caption{Plot of $p_{max}$ as a function of $\theta$}
\label{theta}
\end{figure}

As we know the value of $\theta$, we can easily calculate the value of $\psi_1$ and $\psi_2$ from the above equations and play the local CHSH game for these $\psi_1$ and $\psi_2$. For these values of $\psi_1$ and $\psi_2$, the success probability value corresponding to that $\theta$ will be maximum. Figure~\ref{theta}
shows how $p_{max}$ varies as $\theta$ varies between 0 to $\pi$, taking the maximum value of $cos^2 \pi/8$ at $\theta = \pi/2$.

\subsection{Expected estimation on the sample size}
We recall the Chernoff-Hoeffding~\cite{Chernoff} bound here.
\begin{proposition}
\label{propchernoff}
Let $X=\frac{1}{m}\sum_i{X_i}$ be the average of $m$ independent random variables $X_1, X_2,\cdots,X_m$ with values $[0,1]$, and let $\mathbb{E}[X]=\frac{1}{m}\sum_i{\mathbb{E}[X_i]}$ be the expectation value of $X$, then for any
$\delta>0$, we have $\Pr\left[|X-\mathbb{E}[X]| \geq \delta \right] \leq \exp(-2\delta^2 m).$
\end{proposition}

In our case, if the $i$-th run of the CHSH test succeeds, we set $X_i = 1$; otherwise $X_i = 0$. Note that $\mathbb{E}[X] = \mathbb{E}[X_i] = p$ (say), 
the expected success probability of the CHSH test. The variable $X$ denotes the actual success probability $p'$.

Now the question is how large should ``the number of samples" be so that we get a good ``accuracy" of the given state with  high ``confidence"? More precisely, suppose we want to estimate the success probability $p$ within an error margin of $\epsilon p$ and confidence $1 - \gamma$, meaning 
\begin{equation}
\label{chshsuc}
\Pr[|p' - p| \leq \epsilon p] \geq 1 - \gamma,
\end{equation}
where $p'$ and $p$ are the estimated and the expected values respectively. 
Comparing Equation~\eqref{chshsuc} with Proposition~\ref{propchernoff}, we want, for given $\epsilon$, $p$ and $\gamma$,

\qquad $\exp(-2\epsilon^2 p^2 m) \leq \gamma, \indent \mbox{i.e., }
m \geq \frac{1}{2\epsilon^2 p^2} \ln \frac{1}{\gamma}.$

\begin{figure}[htbp]
\includegraphics[width=0.41\textwidth]{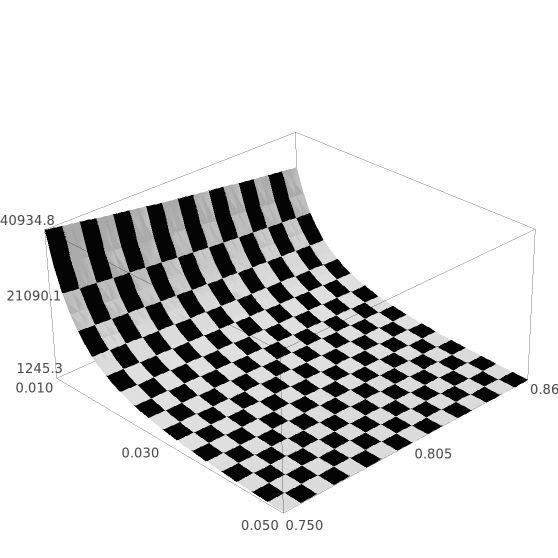}
\caption{Plot of $m_{opt}$ (vertical axis) as a function of $\epsilon$ (left) and $p_{max}$ (right) with $\gamma = 0.01$}
\label{mopt}
\end{figure}

This implies that as the value of the success probability increases, the required sample size decreases. Denoting the maximum success probability for a specific $\theta$ by $p_{max}$, we can write,

$$m_{opt} = \frac{1}{2\epsilon^2 p_{max}^2} \ln \frac{1}{\gamma}.$$
 
This $m_{opt}$ gives the optimal value of the sample size required to certify the states for a given $\theta$. Figure~\ref{mopt} shows how $m_{opt}$ varies with $\epsilon$ and $p_{max}$, when we fix the confidence at 99\%.

\subsection{Security bounds against additional information leakage}
Now, we will propose a bound on the value of $\epsilon$ so that exploiting this deviation, Alice can not extract significant amount of information about the key shared between Bob and herself.
  
In our modified version of the local CHSH test, we suggest that if the maximum success probability of the given state lies within the specified interval then Bob accepts the state and proceeds the protocol otherwise Bob aborts the protocol. 

It may happen that for some other state (for example, $(\alpha |0\rangle_{B} |\phi_0\rangle_{A}+\beta |1\rangle_{B} |\phi_1\rangle_{A})$, where $|\alpha|^2=(\frac{1}{2}+\epsilon_{A})$  and
$|\beta|^2=(\frac{1}{2}-\epsilon_{A})$) the success probability value lies within this interval.  Now to cheat Bob, Alice may supplies a state of the above form. In this case, if Alice chooses the basis 
$\{ |\phi_0\rangle_{A}, |\phi_0^{\perp}\rangle_{A}\}$ with probability $\frac{1}{2}-\epsilon_{A}$ and 
$\{ |\phi_1\rangle_{A}, |\phi_1^{\perp}\rangle_{A}\}$ with probability $\frac{1}{2}+\epsilon_{A}$,
 she can extract $(\frac{1}{2}+2\epsilon_{A}^2)\sin^2\theta$ fraction of entire key stream~\cite{Maitra} which is prohibited by the protocol.
 
 To close such type of security loop-hole (which arises due to the finite sample size) we bound the value of $\epsilon$ so that the additional information which is leaked to Alice should be infinitesimally small.
  
Let, in spite of the claimed state, Bob is provided the states of the form $(\alpha |0\rangle_{B} |\phi_0\rangle_{A}+\beta |1\rangle_{B} |\phi_1\rangle_{A})$. Rigorous calculations show that the success probability for these states merges to $\frac{1}{2} + \frac{1}{8} \sin{\theta} ( \sin{\psi_1} + \sin{\psi_2}) + \frac{1}{4} \sqrt{\frac{1}{4} - \epsilon_{A}^2} (\cos{\psi_1} - \cos{\psi_2}) + \frac{1}{4} \epsilon_{A} \cos{\theta} (\cos{\psi_1} + \cos{\psi_2}) $. We denote this success probability value by $p''$. Now for the given state to be successfully verified, this success probability value ($p''$) must lie within the interval $[ p_{max}-\epsilon p_{max}, p_{max} + \epsilon p_{max} ]$, where $p_{max}$ is the maximum success probability of the original claimed state for a given $\theta$ and 
$\epsilon$ is the accuracy parameter chosen by Bob.

So, $p''$ must satisfy $$p_{max}-\epsilon p_{max} \leq  p'' \leq p_{max}+\epsilon p_{max}.$$ Now from the left and right inequalities, we get
$\epsilon_{A}^{2}\geq -\frac{2\epsilon p_{max}}{\cos{\psi_1}}$
and $\epsilon_{A}^2 \leq \frac{2\epsilon p_{max}}{\cos{\psi_1}}$ respectively.
Since negative $\epsilon_{A}$ is not meaningful, we have the solution as
\begin{equation}
\label{epsaA}
\epsilon_{A}\leq\sqrt{\frac{2\epsilon p_{max}}{\cos{\psi_1}}}.
\end{equation}
Here, we consider only the situation when $\psi_1 \in [0,\frac{\pi}{2})$. This is because from the previous calculation we get that the value of $\psi_1$ always lies within $[0,\frac{\pi}{2})$ whenever $\theta \in [0,\frac{\pi}{2}]$.

So, to deceive Bob the states are prepared in such a way such that the value of $\epsilon_{A}$ must satisfy the condition $\epsilon_{A}\leq\sqrt{\frac{2\epsilon p_{max}}{\cos{\psi_1}}}$. Otherwise, with a high probability the success probability of the given state will not lie within the specified interval and Bob will abort the protocol.

From the earlier section we get that for a given $\theta$, the values of $p_{max}$, $\psi_1$ and $\psi_2$ are fixed. So, we can write $\epsilon_{A}\leq k\sqrt{\epsilon}$, where $k = \sqrt{\frac{2p_{max}}{cos\psi_1}}$ is a constant for a given $\theta$.  In this case, Alice will get the additional information which equals to $\epsilon_{A}^2 \sin^2{\theta}$~\cite{Maitra}. Thus, the information leaked to Alice remains in order of $\epsilon$. If we choose the value of $\epsilon$ sufficiently small, say $10^{-10}$, then we can bound the leakage in the order of $10^{-10}$.

\section{Modified DI-QPQ protocol with optimal samples and Security Analysis}  
Now, we are in the state to propose our modified protocol for finite sample size. 
Bob first calculates the value of $\psi_1$ and $\psi_2$ for which the claimed state attains the maximum success probability. Then from the calculated maximum success probability $p_{max}$, Bob calculates the required optimal sample size $m_{opt}$ for the local CHSH test to certify the states with certain accuracy and confidence. Bob starts with $n = 2m_{opt}$ number of entangled states (see Section~\ref{security} for explanation). Let $\Gamma_{CHSH}$ denote the set which contains the states for local CHSH test, where $|\Gamma_{CHSH}|=m_{opt}$ and $\Gamma_{QPQ}$ denote the set which contains the remaining states, i.e., $|\Gamma_{QPQ}|=n-m_{opt}=m_{opt}$. Bob chooses the states for each of $\Gamma_{CHSH}$ and $\Gamma_{QPQ}$ uniformly at random from the given set of $n$ states. 
Our modified protocol has been described in Algorithm~\ref{algo1}.

Note that the assumptions considered in~\cite{Maitra} also remain valid for our modified version. Explicitly, here also, we assume i) the inherent correctness of the quantum mechanics, ii) no information leakage from the legitimate parties' laboratories, iii) devices are memoryless i.e., each use of the devices is independent and iv) the detectors have unit efficiencies.

\restylealgo{boxed}
\begin{algorithm}[htbp]
{\scriptsize
\begin{enumerate}
 \item For rounds $i \in \{1,\cdots, |\Gamma _{CHSH}|\}$ 

\hspace{5pt} (a) Bob chooses input $x_i\in\{0,1\}$ and $y_i\in\{0,1\}$ uniformly at random. 

\hspace{5pt} (b) If $x_i=0$, he measures the first qubit of the entangled state in $\{|0\rangle, |1\rangle\}$ basis and if $x_i=1$, he measures that in $\{|+\rangle, |-\rangle\}$ basis. 

\hspace{5pt} (c) Similarly, if $y_i=0$, Bob measures the second qubit of the entangled state in $\{|\psi_1\rangle, |\psi_1^{\perp}\rangle\}$ basis and if $y_i=1$, he measures that in $\{|\psi_2\rangle, |\psi_2^{\perp}\rangle\}$ basis, where the values of $\psi_1$ and $\psi_2$ have been calculated previously.\\
\hspace{5pt} (d) The output is recorded as $a_i (b_i) \in\{0,1\}$ for the first and second particle respectively. 
The encoding for $a_i (b_i)$ is performed as follows. 
\begin{itemize}
\item For the first qubit of each pair, if the measurement result is $|0\rangle$ or $|+\rangle$ then $a_i=0$; if the result is $|1\rangle$ or $|-\rangle$ then it would be $1$.
\item For the second qubit of each pair, if the measurement result is $|\psi_1\rangle$ or $|\psi_2\rangle$ then $b_i=0$ ; and if the measurement result is $|\psi_1^{\perp}\rangle$ or $|\psi_2^{\perp}\rangle$, then $b_i=1$.
\end{itemize}

\hspace{5pt} (e) Testing: For the test round $i \in \Gamma_{CHSH}$, define 
  \begin{eqnarray*}
  Y_i=
  \begin{cases}
   1 & \text{if } a_i\oplus b_i=x_i \wedge {y_i}\\
   0 & \text{if } otherwise.
  \end{cases}
 \end{eqnarray*} 
 \item If the value of
 $\frac{1}{|\Gamma_{CHSH}|}\sum_i {Y_i}$ lies within the range $[ p_{max}-\epsilon p_{max} , p_{max}+ \epsilon p_{max} ]$, where $p_{max}$ equals $\frac{1}{8}(\sin{\theta}(\sin\psi_1+\sin\psi_2)+\cos\psi_1-\cos\psi_2)+\frac{1}{2}$ and $\epsilon$ is the accuracy parameter chosen by Bob, Bob proceeds the protocol otherwise Bob aborts the protocol. 
  \item When the local CHSH test at Bob's end is successful, Bob proceeds for the subset $\Gamma_{QPQ}$ and sends one halves of the remaining entangled pairs to Alice.  
 \item Alice performs the private query phase as described in~\cite{Yang}.
 \end{enumerate}
}
\caption{Modified protocol}
\label{algo1}
\end{algorithm}

\subsection{Security Analysis of the Modified Protocol}
\label{security}
The security analysis of the modified protocol follows from the following result (see Appendix A for proof).
\begin{theorem}
\label{thm1}
If for a subset $\Gamma_{CHSH}$ of size $m$, the fraction of the inputs ($x_i$, $y_i$), $i\in\Gamma_{CHSH}$, which satisfy the CHSH condition i.e., $(a_i\oplus b_i=x_i\wedge y_i)$ is equal to $\frac{1}{8}(\sin{\theta}(\sin\psi_1+\sin\psi_2)+\cos\psi_1-\cos\psi_2)+\frac{1}{2}-\delta$, then for the remaining subset $\Gamma_{QPQ}$ of size $n-m$, a fraction of inputs $(x_i,y_i)$, $i\in \Gamma_{QPQ}$, which satisfy the CHSH condition, is also equal to $\frac{1}{8}(\sin{\theta}(\sin\psi_1+\sin\psi_2)+\cos\psi_1-\cos\psi_2)+\frac{1}{2}-\delta$ with a statistical deviation $\nu$.

Here, $\delta=\sqrt{\frac{1}{2m} \ln{\frac{1}{\epsilon_{CHSH}}}}$ and $\nu=\sqrt{\frac{(m+1)}{2(1-\frac{m}{n})m^2}\ln{\frac{1}{\epsilon_{QPQ}}}}$, $\epsilon_{CHSH}$ and $\epsilon_{QPQ}$ are negligibly small value.
\end{theorem}
Essentially, the result means that if the success probability of the local CHSH game for the set $\Gamma_{CHSH}$ varies in the range $[ p_{max}-\epsilon p_{max} , p_{max}+ \epsilon p_{max} ]$, where $p_{max}$ equals $\frac{1}{8}(\sin{\theta}(\sin\psi_1+\sin\psi_2)+\cos\psi_1-\cos\psi_2)+\frac{1}{2}$ and $\epsilon$ is the accuracy parameter, then the success probability of the game for the set $\Gamma_{QPQ}$ would vary in the range $[ p_{max}-\epsilon p_{max} - \nu , p_{max}+ \epsilon p_{max}+ \nu ]$.

Note that in Theorem~\ref{thm1}, if $n$ is close to $m$,
then $\nu$ is no longer guarranteed to be negligible. On the other hand, the choice $n \geq 2m$ makes the coefficient of $\frac{(m+1)}{m^2}\ln \frac{1}{\epsilon_{QPQ}}$ less than $1$ and thus is practically a good choice. 

So far, the entire security analysis, including that of QPQ~\cite{Yang} and DI-QPQ~\cite{Maitra}, is performed under the assumption that the states provided by Alice are all identical. Indeed, when $n$ is infinitely large, Alice cannot have any advantage in non-uniformly biasing the states, as Bob selects the subset $\Gamma_{CHSH}$ uniformly randomly. However, when $n \gg 2m$, but finite, 
 then Alice could inject more bias in the choice of her basis than the threshold $\sqrt{\frac{2\epsilon p_{max}}{\cos{\psi_1}}}$ (from Eq.~\eqref{epsaA}) for a few states and no bias for the remaining states and still she could pass
the CHSH test by Bob. More formally, if she injects a bias $\epsilon'_A$ in $r$ out of $n$ states uniformly at random, then it can be easily shown that to pass the CHSH test, the following condition is required.
\begin{equation}
\label{newepsaA}
\epsilon'_{A}\leq\sqrt{\frac{2n\epsilon p_{max}}{r\cos{\psi_1}}}.
\end{equation}
Thus, by choosing $r \ll n$, Alice can lift the threshold of $\epsilon'_{A}$ much higher than that of $\epsilon_{A}$ and can also retrieve more number of keys if the corresponding states are selected for QPQ. 

To resist this attack, Bob has to choose the minimum possible $n$, i.e., $n=2m$. Since Bob will take $m=m_{opt}$ as per our analysis in Section~\ref{max}, we have $n = 2m_{opt}$.

One may think that the restriction on $n$ would limit Bob to know the key bits for all the positions of the database. This can be easily taken care of by allowing Alice and Bob to play the game repeatedly, each time corresponding to new sets of positions in the database, so as to cover all the positions for Bob.

\section{Discussion and conclusion}
In this current draft, we propose an upgradation of the device independent quantum private query protocol presented by Maitra et al~\cite{Maitra}. We actually modify the protocol for finite sample size. We identify that the suggested protocol in~\cite{Maitra} works perfectly in asymptotic limit. However, for practical implementation we always have to deal with finite sample size. This is why we motivated to upgrade the protocol~\cite{Maitra} for practical purpose. In this regard, we estimate the sample size for the local CHSH test in an optimal way. On other words, we search for a strategy which certify the states with high confidence in such a way that the number of samples required for the testing should be minimal. In this direction, we show that if we deal with the maximum success probability of the game for a given $\theta$, we can reduce the number of samples significantly. We also show that the information leakage to Alice depends on the accuracy parameter $\epsilon$. The order of information leakage is equal to the order of $\epsilon$ meaning the smaller the $\epsilon$ is, the better the security of the protocol.

\section{Appendix A: Lemmas and Proofs}
\begin{lemma}(Serfling~\cite{Serfling})
Let $\{x_1,x_2,\cdots,x_n\}$ be a list of values in $[a,b]$ (not necessarily distinct). Let $\overline{x}=\frac{1}{n}\sum_i x_i$ be the average of these random variables. Let $k$ be the number of random variables $X_1,X_2,\cdots,X_k$ chosen from the list without replacement. Then for any value of $\delta>0$, we have
$\Pr\left[|X-\overline{x}| \geq \delta \right] \leq \exp\left(\frac{-2\delta^2 kn}{(n-k+1)(b-a)}\right),$ where $X=\frac{1}{k}\sum_i X_i$.
\end{lemma}
\begin{lemma}(~\cite{Lim}, Corollary to Serfling Lemma)
Let $\mathbb{X}=\{x_1,x_2...x_n\}$ be a list of (not necessarily distinct) values in $[0,1]$ with the average $\mu_{\mathbb{X}}=\frac{1}{n}\sum_{i=1}x_i$. Let $\mathbb{T}$ be a subset of $\mathbb{X}$ of size $t$ with average $\mu_{\mathbb{T}}=\frac{1}{t}\sum_{i\in\mathbb{T}}x_i$. Let $\mathbb{K}$ be the remaining subset of $\mathbb{X}$ with size $k$ (i.e., $t+k=n$). If the average of the subset $\mathbb{K}$ is $\mu_{\mathbb{K}}=\frac{1}{n-t}\sum_{i\in\mathbb{K}}x_i$, then
for any value of $\epsilon > 0$, we have
$\Pr\left(|\mu_{\mathbb{K}}-\mu_{\mathbb{T}}| \geq \sqrt{\frac{n(t+1)}{2t^2(n-t)}\ln{\frac{1}{\epsilon}}}\right) \leq \epsilon.$
\end{lemma}

\noindent{\bf Proof of Theorem~\ref{thm1}}:
\begin{proof}
We define a random variable $Y_i$ as follows:
  $Y_i= 1$, if $a_i\oplus b_i=x_i \wedge {y_i}$; 0 otherwise.
 Now, we choose a subset $\Gamma_{CHSH}$ of size $m$ and define $Y=\frac{1}{m}\sum_{i\in \Gamma_{CHSH}} Y_i$. Here, $Y$ is called observed average value. Let the expected value of $Y$ for that subset be $\mathbb{E}(Y)=\frac{1}{8}(\sin{\theta}(\sin\psi_1+\sin\psi_2)+\cos\psi_1-\cos\psi_2)+\frac{1}{2}$.  Then applying Chernoff-Hoeffding bound (Proposition~\ref{propchernoff}) we get $\Pr\left[|Y-\mathbb{E}(Y)| \geq \delta \right] \leq \exp(-2 \delta^2 m).$

Let $\epsilon_{CHSH}$ be a negligibly small value. Equating $\exp(-2\delta^2 m)$ with $\epsilon_{CHSH}$ we can find the value of $\delta=\sqrt{\frac{1}{2m} \ln{\frac{1}{\epsilon_{CHSH}}}}$.

 Again, we consider the remaining subset $\Gamma_{QPQ}$ of size $n-m$ and define $Y'=\frac{1}{(n-m)}\sum_{i\in \Gamma_{QPQ}} Y_i$. Now, from Lemma 3, it can be shown that $\Pr(|Y-Y'|\geq \nu) \leq \exp\left(\frac{-2m^2\nu^2(n-m)n}{(m+1)n^2}\right).$

Let $\epsilon_{QPQ}$ be a negligibly small value. Then, equating the R.H.S with $\epsilon_{QPQ}$, we get $\nu$. 
\end{proof}

\end{document}